\documentclass[doublespacing]{elsart}
\usepackage{natbib}
\usepackage{graphicx,color}
\usepackage{amssymb}

\def\prl#1#2#3#4{{ Phys. Rev. Lett.} #1, #2[#3-#4].}
\def\jcp#1#2#3{{ J. Chem. Phys.} #1, #2-#3.}

\def\jpcb#1#2#3{{ J. Phys. Chem. B} #1, #2-#3.}

\begin{document}
\begin{frontmatter}
\title{Intrinsic vulnerabilities to mechanical failure in nanoscale films}
                                                                                
\author[label1]{Pooja Shah},
\author[label1,label2]{Thomas M. Truskett {\corauthref{thanks1}}}
\address[label1]{Department of Chemical Engineering and}
\ead{truskett@che.utexas.edu}
\corauth[thanks1]{Corresponding author. Tel.: +1-512-471-6308; fax:
+1-512-471-7060}
\address[label2]
{Institute for Theoretical Chemistry,
The University of Texas at Austin, Austin, Texas 78712, USA}

\begin{abstract}
We use molecular simulations
to explore how sample dimensions and interfacial properties
impact some generic aspects of the mechanical and structural
behavior of nanoconfined materials.  Specifically, 
we calculate the strain-dependent properties of 
minimum-energy thin-film particle configurations (i.e., inherent structures) 
confined between attractive, parallel substrates.  
We examine how the relationship between the transverse 
strain and the stress tensor (the
equation of state of the energy landscape) 
depends on the properties of the film and substrate.    
We find that both film thickness and film-substrate
attractions influence not only the mechanical properties of thin
films, but also the shape and location of the ``weak spots" where voids 
preferentially form in a film as it is strained beyond its point of 
maximum tensile stress.  The sensitivity of 
weak spots to film properties suggests 
that nanoscale materials may be
intrinsically vulnerabile to specific mechanisms of mechanical
failure. 

\end{abstract}
\end{frontmatter}
\section{Introduction}
Materials confined to very small spatial dimensions behave differently 
than in the bulk. In addition to showing quantum confinement 
effects, they also display thermodynamic, kinetic, and mechanical
limits of stability that depend on sample size, shape, and 
the characteristics of their interfaces.  Specific examples of
property modifications include surface-induced phase transitions,
shifts of the bulk glass transition, and interface-mediated modes of 
mechanical failure
\citep{ggrs99, fv01, hs92}.
Unfortunately, because molecular-scale 
processes in highly inhomogeneous environments are difficult 
to resolve experimentally, a mechanistic picture for 
precisely how nanoconfinement impacts stability has been slow to develop.  
This presents a practical barrier to the 
design of technological applications, in particular those relying 
on solid-state nanostructures to exhibit mechanical 
integrity over a broad range of conditions.

In this Article, we study an elementary 
model system that sheds new 
light on how sample 
dimensions and interfacial properties can influence the mechanical 
behavior of nanoconfined materials.  Specifically, we use
molecular simulations to calculate the strain-dependent properties of 
mechanically-stable films of 
particles confined between attractive, parallel substrates.  
Although analogous studies have been carried out for 
models of isotropic
materials, this is, to our knowledge, the first systematic 
investigation of the relationship between the transverse 
strain and the pressure tensor of the inherent structures
(minimum potential energy configurations) of highly inhomogeneous
films.  Our main finding is that both sample dimensions 
and substrate attractions substantially 
influence not only the mechanical properties of thin films, but also 
the morphology and location of ``weak spots" where voids 
preferentially form in a film as it is strained beyond its point of 
maximum tensile stress.  
Although the precise role that these weak spots 
play in dynamic deformation processes is presently
unknown, they appear intimately linked to material 
failure by quasistatic tensile deformation.  Moreover, 
the sensitivity of weak spots to film properties suggests 
that nanoscale materials may be
intrinsically vulnerabile to specific mechanisms of mechanical
failure. 

Since plastic deformation and failure are inherently 
dynamic events, and since the molecular-scale rearrangements that 
accompany them in amorphous materials are still poorly 
understood, molecular dynamics (MD) simulations would 
appear to represent 
an ideal theoretical tool for their investigation.  In fact, MD 
simulation studies over the past decade have been instrumental 
in gaining insights into deformation processes in polymeric and 
small-molecule materials
~\citep{fl98, gr99, rr01, rr03, stevens01, gersap02, cbr02, vbb04, vp03, yjwnp04}.
These insights have facilitated the interpretation of 
experiments and have aided in the introduction of simple 
theories for viscoplastic flow
~\citep{fl98}.
On the other hand, despite 
recent advancements in algorithms for long-time dynamics, MD 
simulations are still limited to accessing relatively short time and 
length scales.  Thus, it is currently computationally prohibitive to 
use MD to exhaustively explore the effects that sample dimensions and 
interfacial conditions have on the mechanisms of mechanical failure, even 
for simple model systems.  The development of alternative methods 
for probing the molecular-scale origins of failure in glasses is 
still essential.

One complimentary approach is to calculate how the properties 
of a material's mechanically-stable inherent structures depend on the
state of strain.  This strategy has typically been 
implemented in one of 
two ways to investigate deformation, tensile 
strength, and failure of amorphous solids.  The first 
method
~\citep{mas93, has93, ml97, ml98, ml99, ls03a, ls03b, ml04a, ml04b}
calls for subjecting an ensemble of inherent structures, created at a prescribed state
of strain or stress, to an athermal 
quasistatic deformation process consisting of alternating steps of 
small affine deformation followed by potential energy minimization.  
%Malandro and Lacks
~\citet{ml97, ml98, ml99}
have used this protocol 
to investigate the connection 
between strain-induced plastic rearrangements in amorphous materials 
and the annihilation of minima on the material's potential energy 
landscape.  A similar implementation has also 
been used by \citet{ml04a, ml04b}
to study the energy fluctuations
associated with amorphous plasticity and the behavior of elastic 
constants near the onset of material failure.  

Alternatively, one can generate collections of inherent structures 
at each macroscopic strain state of interest by 
mapping equilibrium particle configurations from high-temperature 
simulations to their local potential energy minima \citep{sw82}.  
This procedure, which we adopt here,
has been primarily employed by Debenedetti, Stillinger, and 
collaborators~\citep{sds97, rds99, uds01, shends02} to determine
how the inherent structure pressure of `bulk' glass-formers
depends on density, a relationship that 
has been termed the 
{\em equation of state of an energy landscape} (EOSEL) \citep{dstr99}.  
Several trends have emerged from simulated EOSELs
that give insights into the mechanical
properties of amorphous solids.  For example, inherent structures formed with 
densities higher than the material's Sastry density $\rho_S$ 
(shown in Fig~1a) are structurally homogeneous, 
whereas ``weak spots"~\citep{sds97} 
contained in lower-density equilibrium 
configurations develop into 
fissures or voids upon energy minimization.
Thus, $\rho_S$ is a material property that represents 
the minimum density for which
mechanically-stable solid structures 
can remain structurally homogeneous and void free. 
Moreover, 
the corresponding isotropic tension $-p_{IS}(\rho_S)$ is the 
maximum amount that an inherent structure of that material can sustain 
prior to rupture.  Note that $\rho_S$ obtained from
the EOSEL
is conceptually similar to the density of maximum isotropic tensile 
stress obtained from a quasistatic expansion process, 
and, in fact, recent simulations~\citep{shends02} show quantitative
agreement between these two densities for a number of model materials.
\section{Probing Vulnerabilities to Failure in Nanoscale Films}
The EOSELs of ultrathin films have been recently predicted
by an approximate mean-field theory~\citep{mst04,tg03}, but, 
to our knowledge, they have never been 
determined via molecular 
simulation. One naturally expects far richer behavior in thin films
than in bulk materials because 
interfacial interactions render them 
anisotropic and statistically inhomogeneous.  One consequence, 
illustrated in Fig.~1b, is the emergence of two distinct versions of
the curve shown in Fig.~1a, representing the density dependencies of the
transverse ($p^{\|}_{IS}$) and 
normal ($p^{\bot}_{IS}$) components of the inherent structure pressure tensor. 
This additional dimension of the EOSEL prompts several new questions
pertaining to the film's intrinsic 
vulnerabilities to mechanical failure.
For example, how do film thickness and substrate attractions affect 
the 
directionality of mechanical failure (i.e., which component of the 
stress tensor will exhibit a maximum at a smaller value of strain)?
Moreover, is structural failure (i.e., void formation)
initiated at the strain of maximal stress as it is in
bulk materials~\citep{sds97,dstr99}? If so, do
film thickness and substrate attractions strongly influence the
location and the morphology of the corresponding voids?  

To investigate these issues, we calculated the EOSEL of model films 
comprising $N=864$ Lennard-Jones (LJ) 
particles (truncated and shifted at $r=2.5$~\citep{sds97})
confined between parallel substrates that interact with them
through an effective 9-3 LJ potential,
\begin{equation}
v_{fw}(z)=\frac{2\pi}{3} \rho_w \sigma_w^{3} \epsilon_w 
\left[\frac{2}{15} \left(\frac{\sigma_w}{z}\right)^{9}- 
\left(\frac{\sigma_w}{z}\right)^{3}\right]  
\end{equation}
Here, $z$ is the distance between substrate and particle center, 
$\rho_w \sigma_w^{3}=0.988$, and $\sigma_w =1.0962$~\citep{es77}.
All quantities reported in this work are implicity 
nondimensionalized by the standard combinations of length and
energy scales provided by the effective diameter $\sigma$ and the 
well-depth $\epsilon$ of the LJ film particles, respectively.
Periodic boundary conditions were applied in the $x$ and
$y$ directions to simulate a system of infinite transverse extent.
Additional simulations (not shown here) using $N=1024$ and $N=2048$ 
particles were also carried out to verify that the
results were insensitive to $N$.  

The effects of film thickness 
and strength of the film-substrate attractions were
examined by simulating nine films 
characterized by the permutations of the following 
substrate separations $L \in \{5, 7.5, 10\}$ and 
film-substrate well depths 
$\epsilon_{w} \in \{0.2\epsilon_{fw}, \epsilon_{fw}, 5\epsilon_{fw}\}$,
where $\epsilon_{fw}=1.2771$~\citep{es77}.  The inherent structures required for the
EOSEL analysis were obtained by mapping high-temperature ($T=2.5$) 
equilibrium configurations from a series of fluid 
film densities in the range $\rho=0.2-1.0$ to their local
potential energy minima using LBFGS~\citep{ln89}, a limited-memory quasi-Newton 
multidimensional optimization routine. The equilibrium configurations
were generated by Monte Carlo simulations in the canonical 
ensemble. For every film
type and strain state, 100 equilibrium configurations (separated by 500 Monte
Carlo cycles)
were selected for minimization.  The components of the inherent structure pressure tensor were
calculated both by the global virial equation~\citep{bh54} and by spatially
averaging the local~\citet{ik50} expressions. The two approaches produced
indistinguishable results.

For the structural analysis of
the inherent structure configurations, {\em void space} was 
defined to be the volume of film comprising all points that lie both 
a distance $d>1$ from any particle center 
and $d>(\sigma_w+1)/2$ from either
substrate.  The void probability is then simply the fractional film 
volume available for insertion of an additional 
particle center without creating ``overlap'' with an existing 
particle or substrate.  
For each strain state, the average void probability was calculated by
performing $10^{6}$ trial insertions of hard-core test particles of
unit diameter.  These data were binned according to normal 
position $z$ to
determine the film's inhomogeneous void probability profiles.
The methods described above for both generating and analyzing our
thin-film structures follow directly 
from established techniques for studying bulk
materials, and the interested reader can find more information
in recent review articles 
by~\cite{dstr99,dtls01}. 

As a final point, we discuss the main rationale for choosing
the LJ model for our initial investigation.  First, it is a microscopic 
model that shows qualitatively
realistic mechanical and structural behavior~\citep{sds97}, but
it also is simple enough to allow a systematic
investigation of the effects of
interfaces and confinement on
its properties.
Moreover, since the bulk LJ system is already well 
characterized~\citep{sds97, dstr99}, any confinement-induced deviations
should be relatively straightforward to recognize and interpret.  
Finally, a knowledge of
these deviations will provide a useful basis for understanding 
future simulations on materials with richer molecular attributes.

\section {Results and Discussion}

In this section, we study how the specific properties of our 
model LJ films impact their vulnerability to various
modes of mechanical failure.
To accomplish this, 
we investigate how sample dimensions and film-substrate
attractions affect the strain dependencies of the 
stress tensor and the void-space morphology of 
thin-film inherent structures. 
The locations and shapes of the
voids or `weak spots' that appear in the 
films when 
they are strained beyond their 
maximal stress states provide markers for the corresponding
mechanisms of failure.  We discuss how these mechanisms arise
from structural inhomogeneities in the film 
and the properties of the film-substrate
interfaces. 

Fig.~2a shows the simulated EOSEL of an ultrathin film ($L=5$) 
confined between `weakly attractive' 
substrates ($\epsilon_w = 0.2\epsilon_{fw}$).  
Note that $p^{\bot}_{IS}$
exhibits a minimum at a higher value of 
density than does $p^{\|}_{IS}$, or, 
equivalently, the maximum normal stress corresponds to a
smaller strain than does the maximum transverse stress.  Thus, 
from a mechanical perspective, the inherent structure film can be 
considered more vulnerable to failure in
the normal direction when subjected to plane strain.
Fig.~2a also illustrates that, as is observed in the bulk~\citep{sds97,dstr99}, 
the probability of finding voids in the
film becomes non-neglible for densities below the maximal
stress state.  This connection between mechanical and structural
failure is examined more closely in the void probability profiles and 
film images of Fig.~2b and 2c, respectively.  
As can be seen, the main structural effect of
transverse strain is the formation
of planar voids where the film detaches from one of the substrates, 
indicating an 
{\em adhesive} mechanism for 
mechanical failure.

One natural question to ask is, 
do substrate attractions noticeably affect the film's mechanical and 
structural vulnerabilities to failure?  Fig.~3 shows the EOSEL for
an ultrathin film ($L=5$) with `strongly attractive' substrates 
($\epsilon_w=5\epsilon_{fw}$).
Comparison of Fig.~2a and 3a shows that, 
at least for ultrathin films ($L=5$),
there seem to be some features of the EOSEL that are independent of 
the strength of the substrate attractions.  For example, 
the film confined between strongly 
attractive substrates also shows mechanical vulnerability 
to failure in the normal direction, with $p^{\bot}_{IS}$ 
exhibiting its minimum at a higher value of 
density than $p^{\|}_{IS}$.  Moreover, Fig.~3a shows a
similar correspondence between the attainment of maximal stress 
and 
the initial appearance of voids in the film.
However, the key difference can be found in the location of 
the `weak spots' where strain-induced voids appear.  As 
is evidenced by the void profiles and film images of Fig.~3b and
3c, the strongly attractive substrates promote a
{\em cohesive} rather than adhesive failure mechanism, indicated 
by the formation of planar voids
between the molecular layers in the center of the film.  This mode
of failure results in films with much larger tensile strength and 
toughness than films that fail adhesively.     

The main physical result from Fig.~2 and 3 is that strain-induced 
local stresses in highly inhomogeneous films,
dictated by the relative strengths of the interparticle 
and film-substrate attractions, introduce specific
types of vulnerabilities to failure (e.g., adhesive or cohesive failure).
These vulnerabilities have important implications for the material
properties of the films. 
Interestingly, there is also preliminary evidence to suggest that
such energetic vulnerabilities may strongly
influence the actual mechanisms of dynamic failure events.  
In particular, 
recent MD simulations of polymeric
thin films confined between solid substrates~\citep{gr99} 
have found that the transition from
adhesive to cohesive mechanisms of failure is essentially 
independent of temperature and, moreover, that it is primarily controlled by 
the strength of substrate attractions.  
The fact that both dynamic and EOSEL-type simulation 
studies provide a
similar physical picture is intriguing.  However, more systematic
investigations of the energetic and entropic contributions to 
dynamic failure events in model systems will be
necessary to assess the generality
of these observations.  

To explore the impact of thickness $L$ on the mechanical and
structural properties of nanoscale films, we consider two additional 
cases.  The first,
shown in Fig.~4a, is the EOSEL of a thicker film ($L=10$) confined
between strongly attractive substrates ($\epsilon_w = 5\epsilon_{fw}$).  
Note that considerable asymmetry between $p^{\bot}_{IS}$ and 
$p^{\|}_{IS}$ is still apparent in thicker films, 
with the former component 
attaining a larger magnitude of maximum tensile stress 
at a slightly smaller
value of plane strain.  This behavior reflects the fact that film particles
are strongly attracted to the substrates, making it favorable for the film
to accomodate plane strain by tearing itself
apart internally.  However, the proximity of the
strains of maximal normal and tensile stress in Fig.~4a (as compared
to Fig.~3a) suggests that one can expect
differences between the structural mechanisms of failure for thicker 
versus ultrathin
films.  

To interrogate the microscopic origins of these differences, 
we look to the void 
space analysis. 
As can be seen by comparing Fig.~4b and 4c to 
Fig. 3b and 3c, 
cohesive failure does 
show qualitatively different structural consequences in 
ultrathin and thicker films.  
While ultrathin films fail by forming planar voids between
molecular layers parallel to the substrates, 
thicker films generally fail by forming voids that are `local' and 
more isotropic (more balanced in the transverse and normal directions), 
not so different from the ``bubble-like'' voids that form
from weak spots in bulk materials~\citep{sds97,shends02}.  Note that this
local cohesive failure mechanism produces a film with a much smaller
tensile strength than the ultrathin film of Fig.~3.    

The second thicker-film case ($L=10$) that we examine
is the one confined between neutral 
substrates ($\epsilon_w=\epsilon_{fw}$).  What type
of failure mechanism is expected to prevail for this film?
Fig.~4 shows that in thicker films, like in bulk materials, there is a strong
tendency to cavitate upon expansion due local cohesive stresses.
However, since the substrates are not nearly as attractive in the
neutral wall film, it is not clear at the outset whether 
cavitation will be controlled by 
the cohesive stresses in the normal direction (as in Fig.~4) or in the 
transverse direction.  

The EOSEL shown in Fig.~5a displays
only slight asymmetry in the normal and transverse 
pressure tensor components, indicating that the two aforementioned 
mechanisms for failure (normal cohesive and transverse cohesive) 
are in close competition. In contrast to the results of Fig.~4a,
$p_{IS}^{\|}$ in the neutral wall film exhibits its minimum at a density
slightly larger than does $p_{IS}^{\bot}$.  This suggests that there is 
a preference for the transverse cohesive mechanism for failure.

To see the structural consequences of this type of failure, 
we examine the void probability profiles and the film images of Fig.~5b
and 5c, respectively.  Fig.~5b shows that while voids can appear
anywhere in the film, they show a modestly higher tendency for forming 
near the neutral substrates. 
The film configurations of Fig.~5c confirm that the 
strain-induced voids that form
in the thicker film are also ``bubble-like'' like in Fig.~4c, 
as opposed to the planar voids observed in the ultrathin films of
Fig.~2 and 3.

Finally, we point out there is an additional 
structural consequence of transverse cohesive failure.
Since the initiating voids have a modest preference for
forming near substrates with weak attractions, 
and since they grow in size upon the application
of strain, they can ultimately lead to a secondary
adhesive failure event.  This simply means that it may not be
easy to distinguish between adhesive and transverse cohesive modes
of failure ``after the fact'', i.e., by only examining the 
final structure of the failed sample.
In contrast, we do not find a similar strain-induced 
cascading of failure events for
thin films confined between strongly attractive substrates.

\section{Summary and Conclusions}

To complete our nanoscale film analysis, 
the main features of the EOSELs for all 9 films 
studied are summarized in Table~1.  The information provided
includes the film thickness $L$, the well-depth of the
film-substrate potential $\epsilon_w/\epsilon_{fw}$, 
the direction ($\|$ or $\bot$) of vulnerability to mechanical failure 
(i.e., which stress component shows a maximum at a higher
density), and the corresponding density where this occurs $\rho_S$ (i.e., 
the Sastry density~\citep{sds97,dstr99} for the film).  

The main trends make good physical sense.  For thicker films, 
we see a crossover in vulnerabilities from normal
cohesive failure (when confined between strongly attractive substrates) 
to transverse cohesive failure (when confined between weakly
attractive substrates).  In the former case, strong bonding to the
substrates tears the film apart upon the application of strain, 
whereas the film ruptures under its own strain-induced cohesive 
forces if the substrates are only weakly attractive.
These purely energetic 
trends are in qualititative agreement with the predictions of a 
recently introduced energy landscape-based 
theory for films~\citep{mst04,tg03}. In the thicker film cases
discussed above, 
the weak spots where voids form appear ``bubble-like'', similar to 
bulk material behavior.  
For ultrathin films, we find a crossover in vulnerabilities 
from cohesive failure (when confined between 
strongly attractive substrates) to 
adhesive failure (when confined between weakly attractive substrates).  
Films that fail cohesively have much higher
tensile strength and toughness. These trends are in line
with those seen in recent MD simulations of dynamic failure in polymer
thin-film adhesives~\citep{gr99}.
In the ultrathin film cases, the
weak spots appear planar (parallel to the substrates), and their
location is strongly influenced by the inhomogeneous structural layering
of the films. 

In spite of the intriguing results provided
by the present analysis of
the EOSELs for model thin films, there are many open issues that warrant 
future study.  Even if we confine ourselves to EOSEL thin-film
studies, there remains a need to understand how factors such as substrate
morphology and/or the effects of ``free'' interfaces impact material
stability.  Neither of these issues
are addressed in the present work.  The question of how molecular
connectivity and architecture (from small molecules to polymers)
can impact the mechanical strength, structure, and the 
possible mechanisms for failure is also of great fundamental and practical 
importance.  For instance, would molecules of ellipsoidal shape show 
fundamentally different behavior from the spherically-symmetric
molecules studied here, since the former have been shown~\citep{d04}
to pack much more efficiently than the latter~\citep{ttd00,ttd200}? 
Furthermore, 
how does the present picture change if one considers the EOSEL
under plane stress rather than plane strain conditions?  What about failure
occuring under shear deformations?  Finally, the role that failure mechanisms
obtained via quasistatic analyses play in actual dynamic failure 
processes is still far from resolved, and it is
a issue that demands further scrutiny.

\noindent {\bf Acknowledgements}

It is a pleasure to present this work about
mechanical properties, structural inhomogeneities, and void-space 
geometry in nanoscale films for an issue honoring Salvatore
Torquato. He is a lifelong mentor to one of us (TMT) 
and has contributed greatly to
the current state of knowledge on these fascinating topics.
TMT gratefully acknowledges the financial support of NSF (CAREER
CTS-0448721), the David and Lucile Packard Foundation, and the
Donors of the American Chemical Society Petroleum Research Fund.

\pagebreak
\noindent {\bf Figure Captions}
\begin{enumerate}
\item Schematic of the relationship between inherent structure (IS) pressure and
density, the so-called equation of state of an energy landscape (EOSEL). (a) In 
bulk isotropic materials, inherent structures formed at densities above the Sastry density
$\rho_S$ are structurally homogeneous. Those formed at densities below $\rho_S$,
the state of maximum isotropic tensile stress, contain voids or fissures (Sastry
et al., 1997; Debenedetti et al., 1999). (b) Schematic of the EOSEL for a thin
film, showing the diagonal components of the inherent structure pressure tensor in the
transverse ($p^{\|}_{IS}$) and normal ($p^{\bot}_{IS}$) directions. \\

\item(a) EOSEL of an ultrathin film ($L=5$) confined between weakly attractive
substrates ($\epsilon_w = 0.2 \epsilon_{fw}$), depicting normal ($\Box$) and
transverse ($\triangleright$) components of the inherent structure pressure tensor. Also shown
is the corresponding probability of finding of voids ($\circ$) in the film.
(b) Void probability distribution $P(z)$ in the normal direction for
$\rho = 0.2, 0.65, 0.66,$ and $1.0$. The arrow indicates increasing density
(decreasing transverse strain). (c) Visual depiction of inherent structure configurations for
densities $\rho=0.5$
(left) and $\rho=0.66$ (right). These illustrate where planar voids form near
the substrate, indicating an adhesive failure mechanism. \\

\item(a) EOSEL of an ultrathin film ($L=5$) confined between strongly attractive
substrates ($\epsilon_w = 5 \epsilon_{fw}$), depicting normal ($\Box$) and
transverse ($\triangleright$) components of the inherent structure pressure tensor. Also shown
is the corresponding probability of finding of voids ($\circ$) in the film.
(b) Void probability distribution $P(z)$ in the normal direction for
$\rho = 0.2, 0.3, 0.76,$ and $1.0$. The arrow indicates increasing density
(decreasing transverse strain). (c) Visual depiction of inherent structure configurations for 
densities $\rho=0.5$
(left) and $\rho=0.76$ (right). These illustrate where planar voids form 
in between molecular layers, indicating a cohesive failure mechanism. \\

\item(a) EOSEL of a thicker film ($L=10$) confined between strongly attractive
substrates ($\epsilon_w = 5 \epsilon_{fw}$), depicting normal ($\Box$) and
transverse ($\triangleright$) components of the inherent structure pressure tensor. Also shown
is the corresponding probability of finding of voids ($\circ$) in the film.
(b) Void probability distribution $P(z)$ in the normal direction for
$\rho = 0.2, 0.86, 0.87,$ and $1.0$. The arrow indicates increasing density
(decreasing transverse strain). (c) Visual depiction of inherent structure configurations for densities $\rho=0.6$
(left) and $\rho=0.82$ (right). These illustrate 
where ``bubble-like'' voids form in the film, consistent with 
a normal cohesive
failure mechanism. \\

\item(a) EOSEL of a thicker film ($L=10$) confined between neutral 
substrates ($\epsilon_w = \epsilon_{fw}$), depicting normal ($\Box$) and
transverse ($\triangleright$) components of the inherent structure pressure tensor. Also shown
is the corresponding probability of finding of voids ($\circ$) in the film.
(b) Void probability distribution $P(z)$ in the normal direction for
$\rho = 0.2, 0.81, 0.82,$ and $1.0$. The arrow indicates increasing density
(decreasing transverse strain). (c) Visual depiction of inherent
structure 
configurations for
densities $\rho=0.4$
(left) and $\rho=0.82$ (right). These illustrate where ``bubble-like'' voids 
form near the substrates, consistent with 
a transverse cohesive failure mechanism. \\

\end{enumerate}
\newpage
\begin{figure}
\centerline{\includegraphics[scale=0.7]{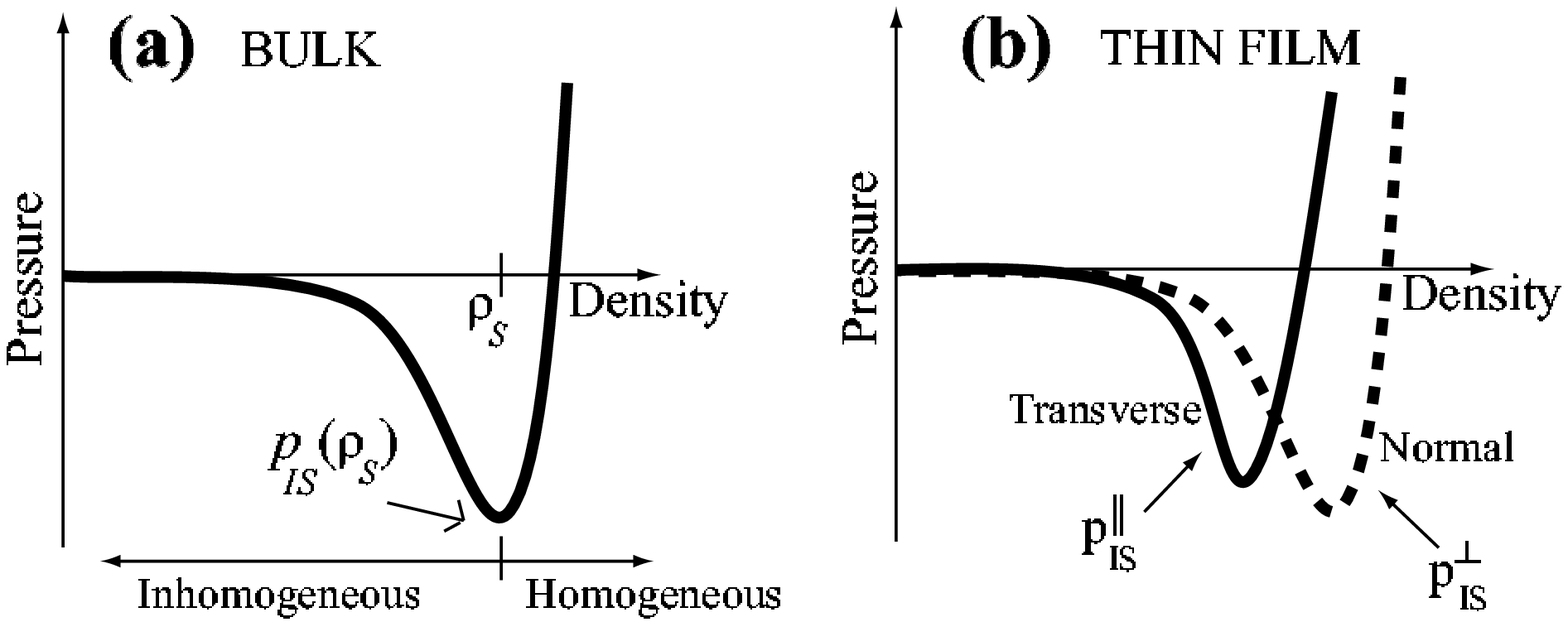}}
\caption{}
\end{figure}
\newpage
\newpage
\begin{figure}
\centerline{\includegraphics[scale=0.7]{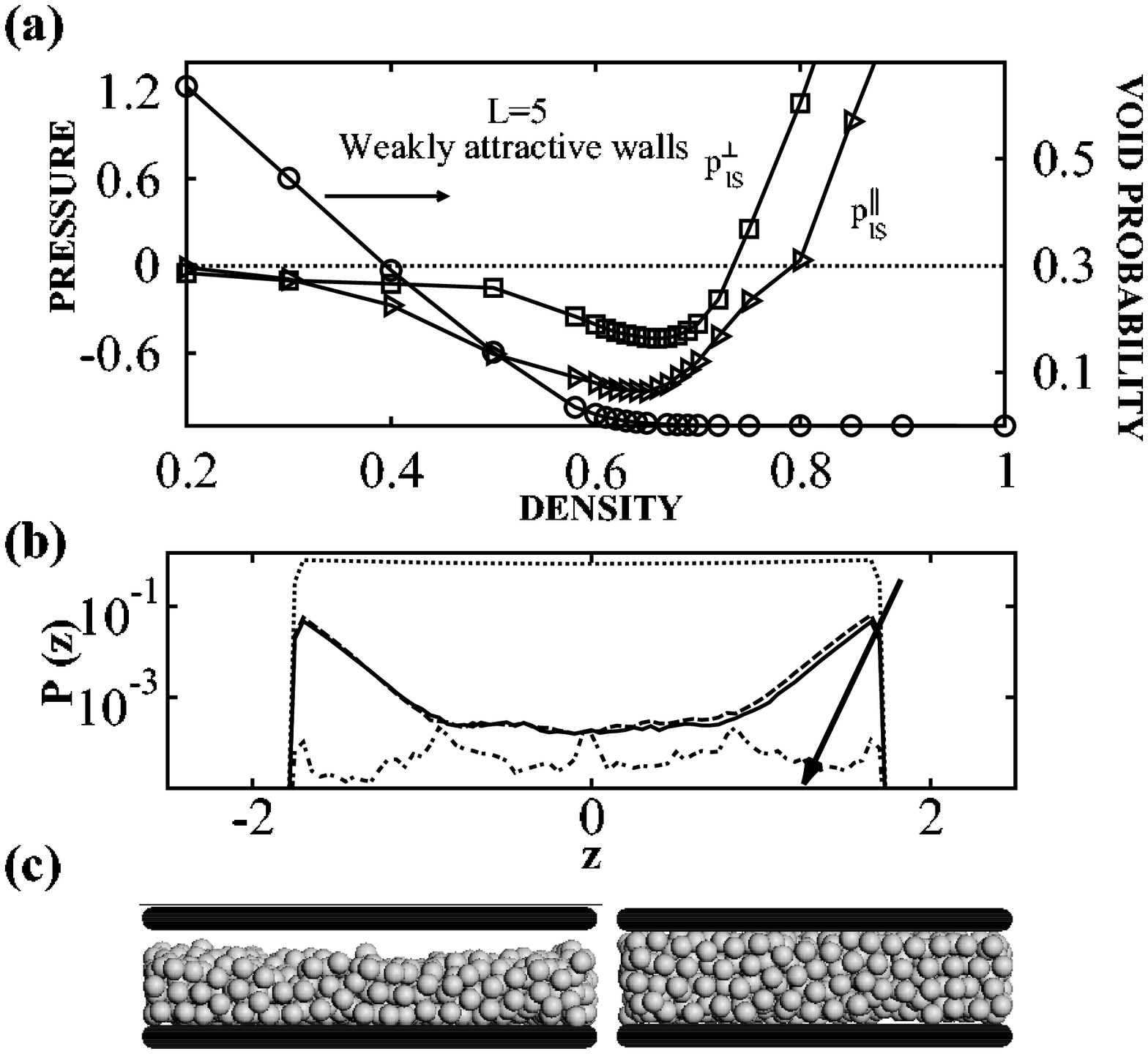}}
\caption{}
\end{figure}
\newpage
\begin{figure}
\centerline{\includegraphics[scale=0.7]{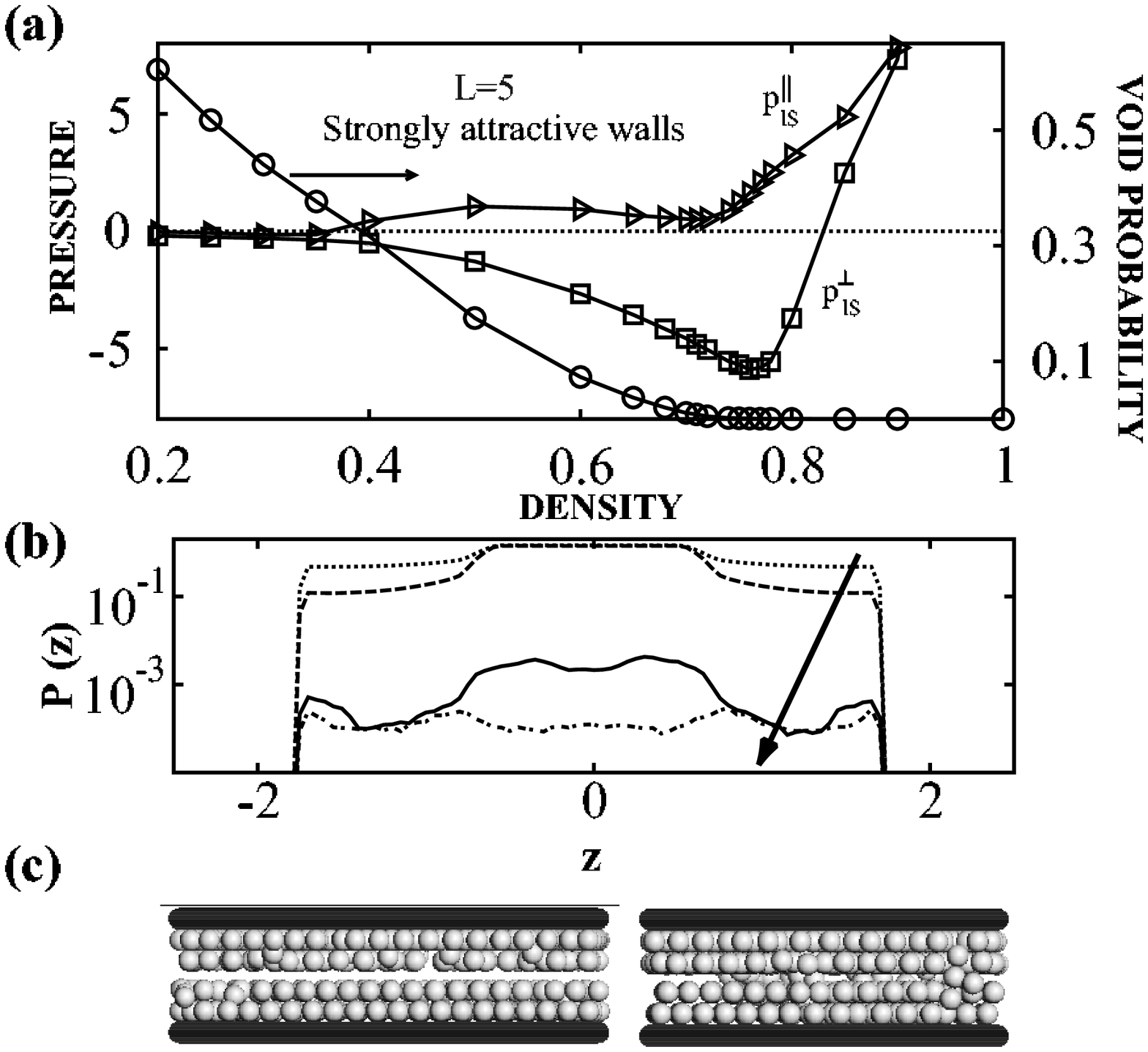}}
\caption{}
\end{figure}
\newpage
\begin{figure}
\centerline{\includegraphics[scale=0.7]{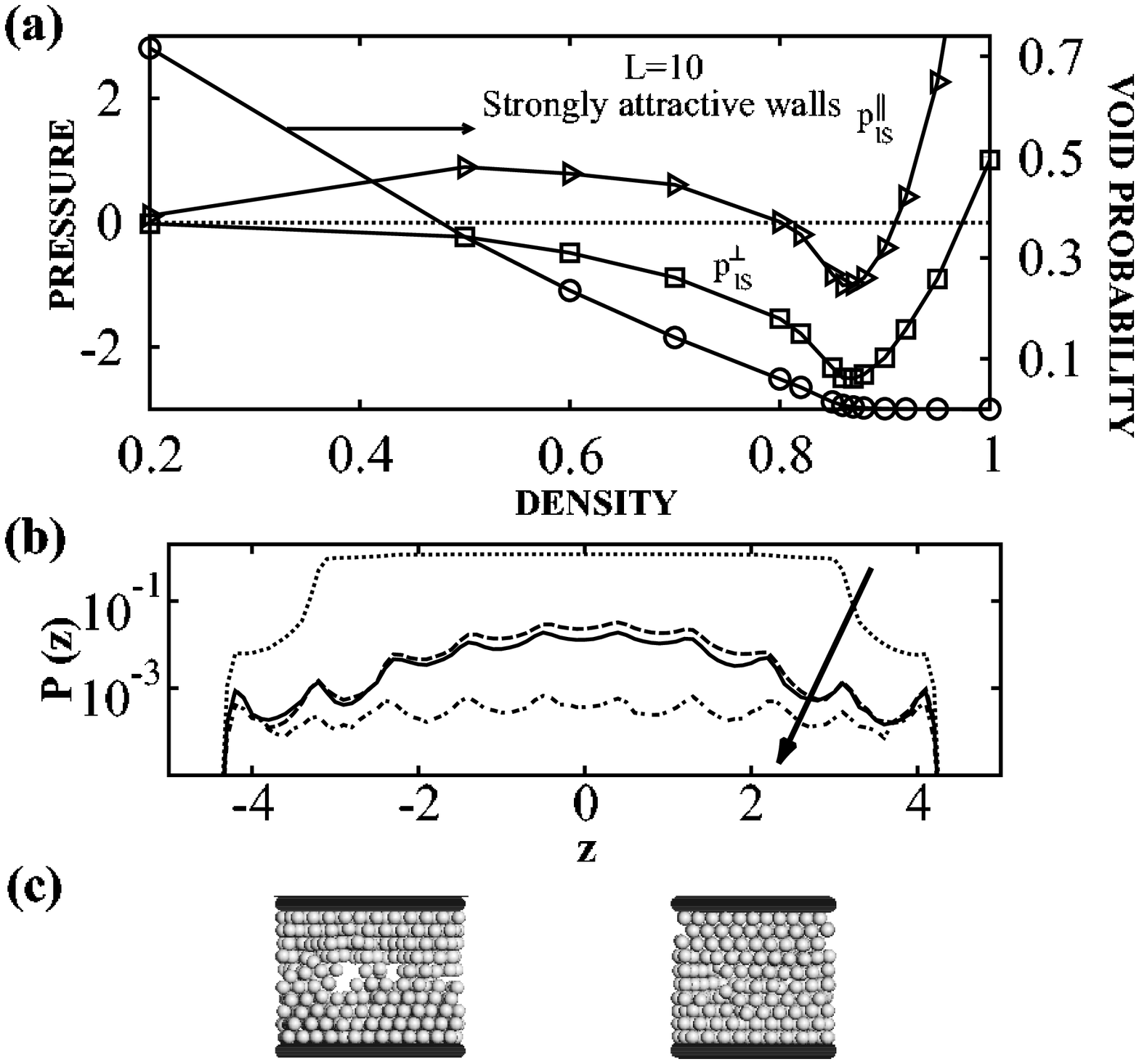}}
\caption{}
\end{figure}
\newpage
\begin{figure}
\centerline{\includegraphics[scale=0.7]{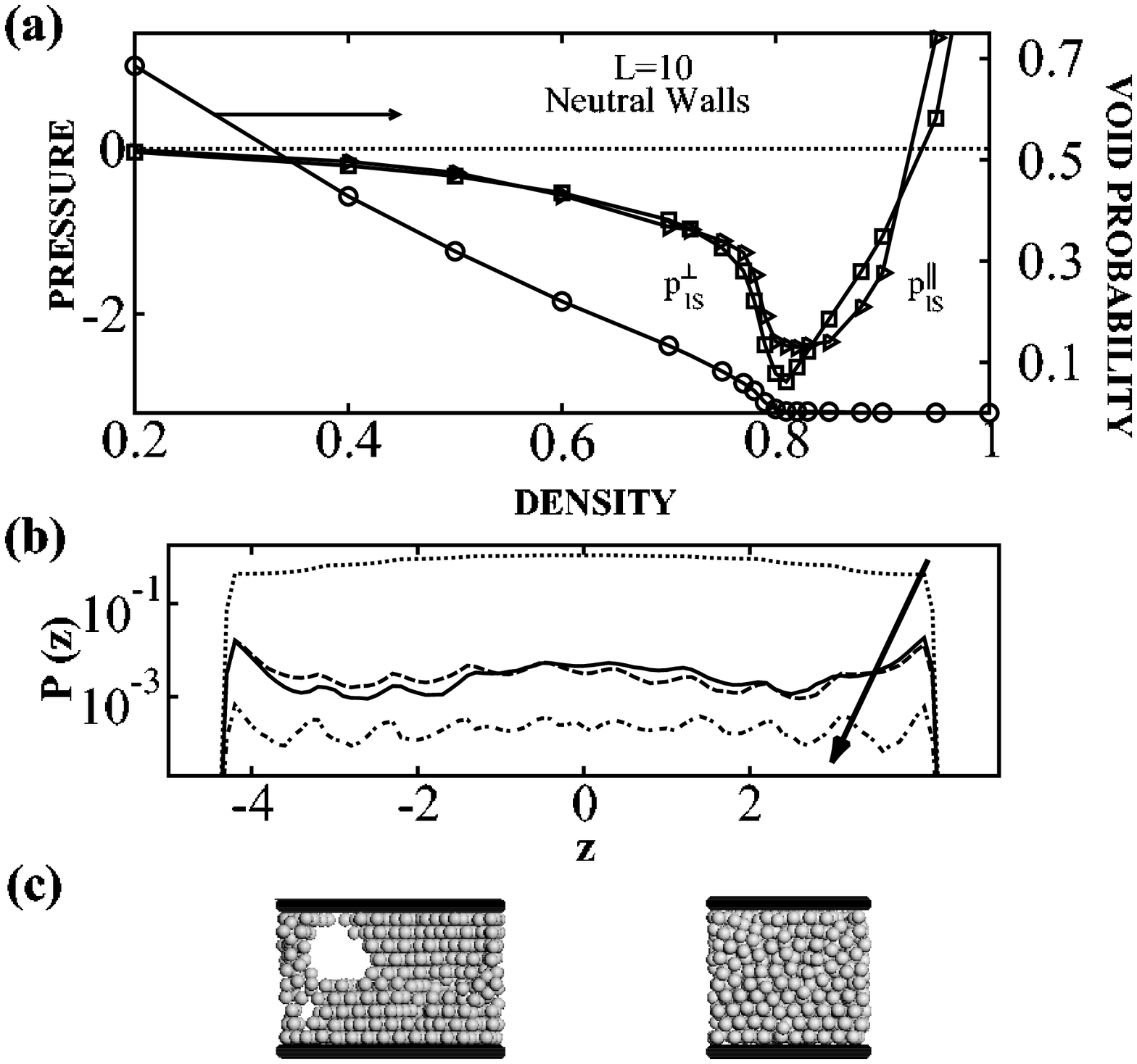}}
\caption{}
\end{figure}
\newpage
\begin{center}
\begin{table}
\begin{center}
\setlength{\tabcolsep}{6mm}
\begin{tabular}{|c|c|c|c|c|}
\hline
$L$ & $\epsilon_w/\epsilon_{fw}$ & Failure direction & $\rho_S$ \\
\hline
& 5.0 & $\bot$ & 0.87 \\
10 & 1.0 & $\|$ & 0.82 \\ 
& 0.2 & $\|$ & 0.82 \\
\hline
 & 5.0 & $\bot$ & 0.85 \\ 
7.5 & 1.0 & $\bot$ & 0.81 \\ 
& 0.2 & $\|$ & 0.77 \\
\hline
 & 5.0 & $\bot$ & 0.76 \\ 
5 & 1.0 & $\bot$ & 0.70 \\ 
& 0.2 & $\bot$ & 0.66\\ 
\hline
\end {tabular}
\end{center}
\caption{
%\small
{Summary of the main features of the EOSELs for 9 films: film thickness $L$, the
well-depth of the film-substrate potential $\epsilon_w/\epsilon_{fw}$,
the primary direction ($\|$ or $\bot$) of vulnerability to mechanical failure,
and the corresponding density where this occurs $\rho_S$ (i.e., the Sastry
density for the film). 
}} 
\end{table}
\end{center}

\begin{thebibliography}{}

\bibitem
[Born and Huang, 1954]
{bh54} 
Born, M., Huang, K., 1954. 
Dynamical Theory of Crystal Lattices, Claredon Press, Oxford.

\bibitem
[Capaldi et al., 2002]
{cbr02} 
Capaldi, F.~M., Boyce, M.~C., Rutledge, G.~C., 2002. Enhanced 
mobility accompanies the active deformation of a glassy amorphous polymer. 
Phys. Rev. Lett. 89, 175505[1-4]. 
%Physics Review Letters, 89, 175505. 

\bibitem
[Debenedetti et al., 1999]
{dstr99} 
Debenedetti, P.~G., Stillinger, F.~H., Truskett, T.~M., 
Roberts, C.~J., 1999.
The equation of state of an energy landscape.
\jpcb{103}{7390}{7397}
%J. Phys. Chem. B {\bf 103}, 7390 (1999).

\bibitem
[Debenedetti et al., 2001]
{dtls01}
Debenedetti, P.~G., Truskett, T.~M., Lewis, C.~P., Stillinger, F.~H., 2001.
Theory of supercooled liquids and glasses:  Energy landscape and
statistical geometry perspectives.
Adv. Chem. Eng. 28, 21-79.

\bibitem
[Donev et al., 2004]
{d04} 
Donev, A., Cisse, I., Sachs, D., Variano, A. Stillinger,
F.~H. Connelly, R., Torquato, S. and Chaikin, P.~M., 2004.
Improving the density of jammed disordered packings using elipsoids. 
Science 303, 990-993.

\bibitem
[Ebner and Saam, 1977]
{es77} 
Ebner, C., Saam, W.~F., 1977.
New phase-transition phenomena in thin argon films.
Phys. Rev. Lett. 38, 1486-1489.
%\prl{38}{1486}{1489}

\bibitem
[Falk and Langer, 1998]
{fl98} 
Falk, M.~L., Langer, J.~S., 1998. Dynamics of viscoplastic 
deformation in amorphous solids. 
Phys. Rev. E 57, 7192-7205.

\bibitem
[Forrest and Dalnoki-Veress, 2001]
{fv01} 
Forrest, J.~A., Dalnoki-Veress, K., 2001. The glass transition
in thin polymer films. 
Adv. Colloid Interface Sci. 94, 167-195. 

\bibitem
[Gelb et. al, 1999]
{ggrs99}
Gelb, L.~D., Gubbins, K.~E., Radhakrishnan, R., Sliwinska-Bartkowiak, M., 1999.
Phase separation in confined systems. 
Rep. Prog. Phys. 62, 1573-1659.

\bibitem
[Gersappe and Robbins, 1999]
{gr99} 
Gersappe, D., Robbins, M.~O., 1999. Where do polymer adhesives fail?. 
Europhys. Lett. 48, 150-155.
%Europhysics Letters, 48, 150-155.

\bibitem
[Gersappe, 2002]
{gersap02} 
Gersappe, D., 2002. 
Molecular mechanisms of failure in polymer nanocomposites. 
Phys. Rev. Lett. 89, 058301[1-4].
%Physics Review Letters, 89, 058301.
 
\bibitem
[Hutchinson and Suo, 1992]
{hs92} 
Hutchinson, J.~W., Suo, Z., 1992. Mixed mode cracking in layered
materials, in: Hutchinson, J.~W., Wu, T.~Y. (Eds.), Advances in Applied 
Mechanics. Academic Press, San Diego, pp. 63-191.

\bibitem
[Hutnik et al., 1993]
{has93} 
Hutnik, M., Argon, A.~S., Suter, U.~W., 1993. 
Simulation of elastic and plastic response in the glassy polycarbonate of 
4,4-isopropylidenediphenol. 
Macromolecules 26, 1097-108.

\bibitem
[Irving and Kirkwood, 1950]
{ik50} 
Irving, J.~H., Kirkwood, J.~G., 1950. 
The statistical mechanical theory of transport processes. IV. The equation of
hydrodynamics. 
\jcp{18}{817}{829}
%J. Chem. Phys. {\bf 18}, 817 (1950). 

\bibitem
[Liu and Nocedal, 1989]
{ln89} 
Liu, D.~C., Nocedal, J., 1989.
On the limited memory method for large scale optimization.
Math. Program. B 45, 503-528. 

\bibitem
[Lund and Schuh, 2003a]
{ls03a} Lund, A.~C., Schuh, C.~A., 2003a. 
Yield surface of a simulated metallic glass.
Acta Mater. 51, 5399-5411.
%Acta Materials, 51, 5399.

\bibitem
[Lund and Schuh, 2003b]
{ls03b} 
Lund, A.~C., Schuh, C.~A., 2003b.
Driven alloys in the athermal limit.
\prl{91}{235505}{1}{4}
%Physical Review Letters, 91, 235505.

\bibitem
[Malandro and Lacks, 1997]
{ml97} 
Malandro, D.~L., Lacks, D.~J., 1997.
Volume dependence of potential energy landscapes in glasses.
\jcp{107}{5804}{5810} 
%Journal of  Chemical Physics, 107, 5804-5810. 

\bibitem
[Malandro and Lacks, 1998]
{ml98} 
Malandro, D.~L., Lacks, D.~J., 1998.
Molecular-level mechanical instabilities and enhanced self-diffusion in flowing 
liquids. 
Phys. Rev. Lett. 81, 5576-5579.
%\prl{81}{5576}{5579} 
%Physics Review Letters, 81, 5576-5579. 

\bibitem
[Malandro and Lacks, 1999]
{ml99} 
Malandro, D.~L., Lacks, D.~J., 1999.   
Relationships of shear-induced changes in the potential energy landscape to the 
mechanical properties of ductile glasses.
\jcp{110}{4593}{4601}
%Journal of  Chemical Physics, 110, 4593-4601.

\bibitem
[Maloney and Lema\^{i}tre, 2004a] 
{ml04a} 
Maloney C., Lema\^{i}tre, A., 2004a. 
Subextensive scaling in the athermal, quasistatic limit of amorphous matter in 
plastic shear flow.
\prl{93}{016001}{1}{4}
%Phys. Rev. Lett. {\bf93}, 016001 (2004); 

\bibitem
[Maloney and Lema\^{i}tre, 2004b] 
{ml04b} 
Maloney C., Lema\^{i}tre, A., 2004b. 
Universal breakdown of elasticity at the onset of material failure.
\prl{93}{195501}{1}{4}

\bibitem
[Mittal et al., 2004]
{mst04} 
Mittal, J., Shah, P., Truskett, T.~M., 2004.
Using energy landscapes to predict the properties of thin films.
J. Phys. Chem. B, 108, 19769-19779.

\bibitem
[Mott et al., 1993]
{mas93} 
Mott, P.~H., Argon, A.~S., Suter, U.~W., 1993.
Atomistic modelling of plastic deformation of glassy polymers.
Phil. Mag. A 67, 931-978.
%Philosophical Magazine A, 67, 931-978.

\bibitem
[Roberts et al., 1999]
{rds99} 
Roberts, C.~J., Debenedetti, P.~G., Stillinger, F.~H., 1999.
Equation of state of the energy landscape of SPC/E water. 
\jpcb{46}{10258}{10265}
%J. Phys. Chem. B {\bf 46}, 10258 (1999); 

\bibitem
[Rottler and Robbins, 2001]
{rr01} 
Rottler, J., Robbins, M.~O., 2001. Yield conditions for 
deformation of amorphous polymer glasses. 
Phys. Rev. E 64, 051801[1-8].
%Physical Review E, 64, 051801.

\bibitem
[Rottler and Robbins, 2003]
{rr03} 
Rottler, J., Robbins, M.~O., 2003. Shear yielding of amorphous 
glassy solids: Effect of temperature and strain rate. 
Phys. Rev. E 68, 011507[1-10]. 
%Physical Review E, 68, 011507. 

\bibitem
[Sastry et al., 1997]
{sds97} 
Sastry, S., Debenedetti, P.~G., Stillinger, F.~H., 1997. 
Statistical geometry of particle packings. II. ``Weak spots" in liquids.
Phys. Rev. E 56, 5533-5543. 

\bibitem
[Shen et al., 2002]
{shends02} 
Shen, V.~K., Debenedetti, P.~G., Stillinger, F.~H., 2002.
Energy landscape and isotropic tensile strength of n-alkane glasses.
\jpcb{106}{10447}{10459} 
%  J. Phys. Chem. B {\bf 106}, 10447 (2002). 

\bibitem
[Stevens, 2001]
{stevens01} 
Stevens, M.~J., 2001. 
Interfacial fracture between highly cross-linked polymer networks and a solid 
surface: Effect of interfacial bond density. 
Macromolecules 34, 2710-2718.

\bibitem[Stillinger and Weber, 1982]{sw82} Stillinger, F.~H., Weber, T.~A.,
1982. Hidden structures in liquids. Phys. Rev. A 25, 978-989.

\bibitem[Torquato et al., 2000]{ttd00} Torquato, S., Truskett, T.~M.,
  Debenedetti, P.~G., 2000.  Is random close packing of spheres well
  defined? Phys. Rev. Lett. 84, 2064-2067.

\bibitem[Truskett and Ganesan, 2003]{tg03} Truskett, T.~M., Ganesan, V.,
2003. Ideal glass transitions in thin films. J. Chem. Phys. 119, 1897-1900.

\bibitem[Truskett et al., 2000]{ttd200} Truskett, T.~M., Torquato, S.,
  Debenedetti, P.~G., 2000.  Towards a quantification of disorder in
  materials:  Distinguishing equilibrium and glassy sphere packings. Phys. Rev. E 62, 993-1001.


\bibitem
[Utz et al., 2001]
{uds01} 
Utz, M., Debenedetti, P.~G., Stillinger, F.~H., 2001.
Isotropic tensile strength of molecular glasses.
\jcp{114}{10049}{10057}
%J. Chem. Phys. {\bf 114}, 10049 (2001).

\bibitem
[Van Workum and de Pablo, 2003]
{vp03} 
Van Workum, K., de Pablo, J. J., 2003. 
Computer simulation of the mechanical properties of amorphous polymer 
nanostructures. 
Nano Lett. 3, 1405-1410. 
%Nano Letters, 3, 1405-1410. 

\bibitem
[Varnik et al., 2004]
{vbb04} 
Varnik, F., Bocquet, L., Barrat, J.-L., 2004. 
A study of the static yield stress in a binary Lennard-Jones glass. 
\jcp{120}{2788}{2801}
%Journal of Chemical Physics, 120, 2788-2801.

\bibitem
[Yoshimoto et al., 2004]
{yjwnp04} 
Yoshimoto, K., Jain, T.~S., Van Workum, K., Nealey, P.~F., 
de Pablo, J.~J., 2004. 
Mechanical heterogeneities in model polymer glasses at small length scales. 
\prl{93}{175501}{1}{4}
%Physics Review Letters, 93, 175501.
\end{thebibliography}
\end{document}